# High-$T_c$ superconductivity in ultrathin Bi$_2$Sr$_2$CaCu$_2$O$_{8+x}$ down to half-unit-cell thickness by protection with graphene

Da Jiang[1], Tao Hu[1], Lixing You[1], Qiao Li[1], Ang Li[1], Haomin Wang[1], Gang Mu[1], Zhiying Chen[1], Haoran Zhang[1], Guanghui Yu[1], Jie Zhu[2], Qiujuan Sun[1,3], Chengtian Lin[4], Hong Xiao[5], Xiaoming Xie[1,6], Mianheng Jiang[1,6]

[1]State Key Laboratory of Functional Materials for Informatics, Shanghai Institute of Microsystem and Information Technology, Chinese Academy of Sciences, 865 Changning Road, Shanghai 200050, China.

[2]Insitiute of Precision Optical Engineering, School of Physics science and Engineering, Tongji University, Shanghai 200092, China.

[3]School of Physics and Electronics, Central South University, Changsha 410083, China.

[4]Max-Planck-Institute für Festkörperforschung, 70569 Stuttgart, Germany.

[5]Beijing National Laboratory for Condensed Matter Physics, Institute of Physics, Chinese Academy of Sciences, Beijing 100190, China.

[6]School of Physical Science and Technology, ShanghaiTech University, 319 Yueyang Road, Shanghai 200031, China.

Correspondence and requests for materials should be addressed to D.J. (jiangda@mail.sim.ac.cn) or T.H. (thu@mail.sim.ac.cn).

**Abstract**

High-$T_c$ superconductors confined to two dimension exhibit novel physical phenomena, such as superconductor-insulator transition. In the $Bi_2Sr_2CaCu_2O_{8+x}$ (Bi2212) model system, despite extensive studies, the intrinsic superconducting properties at the thinness limit have been difficult to determine. Here we report a method to fabricate high quality single-crystal Bi2212 films down to half-unit-cell thickness in the form of graphene/Bi2212 van der Waals heterostructure, in which sharp superconducting transitions are observed. The heterostructure also exhibits a nonlinear current-voltage characteristic due to the Dirac nature of the graphene band structure. More interestingly, although the critical temperature remains essentially the same with reduced thickness of Bi2212, the slope of the normal state $T$-linear resistivity varies by a factor of 4-5, and the sheet resistance increases by three orders of magnitude, indicating a surprising decoupling of the normal state resistance and superconductivity. The developed technique is versatile, applicable to investigate other two-dimensional (2D) superconducting materials.

**Introduction**

Two-dimensional superconductors provide insights into the mechanisms of superconductivity and their possible device applications[1-11]. Bi2212 is a quasi-2D cuprate in which thin superconducting sheets ($CuO_2$) alternate with insulating layers (SrO, BiO). Its extreme anisotropy[12-15] makes Bi2212 an ideal material to address the fundamental question whether superconductivity truly survives down to half-unit-cell thickness which may in turn help to answer another fundamental question: what the superconducting mechanisms are in the ultrathin Bi2212. The superconducting properties of ultrathin Bi2212 have already been studied in both bottom-up[4,5] (thin-film growth) and top-down[6] (micro-fabrication of single crystal) approaches. Unfortunately, these approaches have yielded inconsistent results, with behavior ranging from insulating to superconducting. Molecular beam epitaxy (MBE) method has been used to grow half-unit-cell-thick Bi2212 (HUC-Bi2212) sandwiched in between multiple Bi2201 blocks[8,9]. In these ultrathin MBE superlattices, the resulting superconducting properties depend on the sandwich structure[16-19], which results in uncertainty about the properties originate from HUC-Bi2212 or from the superlattice stack. Inspired by the successful isolation of graphene by micromechanical exfoliation[20], researchers have prepared many 2D materials of extremely high quality using this unique and versatile technique, including $MoS_2$, $NbSe_2$, and single-unit-cell-thick Bi2212 (SUC-Bi2212), which was shown to be insulating[7]. Recently, the use of a van der Waals heterostructure composed of graphene and other 2D atomic crystals has emerged as a promising approach to pursue condensed matter physics questions in these systems[21], such as Hofstadter's butterfly effect and the fractal quantum Hall effect. This structure, which is artificially constructed by stacking different 2D crystals on top of each other, exhibits many unusual properties[21,22], such as quantum capacitance and moiré patterns.

Here we use mechanical exfoliation method to study the intrinsic properties of Bi2212, from half-unit-cell thickness to bulk material, by producing thin films of graphene/Bi2212 van der Waals heterostructures. In addition to superconducting transition above the liquid nitrogen temperature observed in all the samples, we were surprised to find the decoupling of the normal state resistance and superconductivity, which contrasts with previous reports[23] on their importance to the superconducting mechanism.

## Results

**Temperature dependence of resistance in graphene/SUC-Bi2212**

Samples were prepared by mechanical exfoliation method (Fig. 1). Four-probe measurements were performed on graphene/Bi2212 heterostructure samples at the temperature ($T$) from 2 to 300 K by using direct current (dc) combined with alternating current (ac) ($I_{dc}+I_{ac}$). Superconductivity was observed in all samples even at half-unit-cell thickness, although the success rate for the preparation of the HUC2212 heterostructure was lower.

We observed a sharp resistive transition above the liquid nitrogen temperature (at 88 K) regardless of $I_{dc}$ bias in the 0 to 0.36 mA range examined (Fig. 2a), indicating that robust superconductivity is preserved. In the heterostructure, graphene and SUC-Bi2212 form a parallel circuit as illustrated in the inset to Fig. 2a. We also observed residual resistance in the superconducting state, as a result of partial current flowing through the graphene protection layer (although we eliminated this by converting the graphene to an insulator, as discussed in the next subsection). For small bias currents (less than 0.12 mA in our case) in the superconducting temperature range, the differential resistance increased with decreasing temperature, which suggests opening of a superconducting gap. In contrast, the differential resistance in the superconducting state became independent of temperature at high bias current, which overcomes the gap. After subtracting the temperature independent resistance component from the data, all the curves precisely overlap with one another (Fig. 2b), indicating the consistency and reliability of our measurements and the reproducibility of our fabrication technique. Furthermore, different samples with the same thickness can be scaled with each other (Fig. 2c). The *R-T* curves in Fig. 2b thus reflect the intrinsic superconducting properties of SUC-Bi2212.

We also observed a *T* independent parabolic-like fixed offset *I-V* curve in the superconducting state of the graphene/SUC-Bi2212 heterostructure (Fig. 2d). This non-vanished voltage originates from the two junctions between SUC-Bi2212 and graphene as illustrated in the inset of Fig. 2a. Graphene is known to be a Dirac fermion material with linear energy dispersion at the Dirac point[20]. The differential conductance of the junction *dI/dV* is a result of tunneling from graphene to SUC-Bi2212. The value is thus proportional to the product of the density of states of graphene ($n_{graphene}$) and SUC-Bi2212 ($n_{SUC-Bi2212}$), which is $dI/dV \propto n_{SUC-Bi2212} \times n_{graphene}$ (Ref. 26). The value of $n_{SUC-Bi2212}$ is not affected by the voltage in the superconducting state because there is zero electric field inside the superconductor, however, $n_{graphene}$ is proportional to the bias voltage ($V_g$) across the junction ($V_g \propto n_{graphene}$) (Ref. 27). When performing *I-V* measurements below $T_c$, the voltage (*V*) drops only at the graphene/SUC-Bi2212 junctions, which is $V=2V_g$. Therefore, we obtain $dI/dV \propto V$, hence, $I \propto V^2$, which is precisely the parabolic offset of the *I-V* curves presented in Fig. 2d. The voltage actually drives the single layer graphene away from the Dirac point and tunes its density of states, which yields a parabolic *I-V* curve.

**The role of graphene in the heterostructure**

SUC-Bi2212 is known to be insulating[6,7]. The present investigation confirms that ultrathin Bi2212 transforms into an insulator if graphene is not transferred onto it. This observation suggests that one of the major roles of graphene is to protect the underlying ultrathin Bi2212, because graphene is highly impermeable even to small atoms such as helium[28].

To verify that the residual resistance is a result of the current bypass effect, we converted the graphene capping layer into an insulator by oxidation[29]. Interestingly, this caused the SUC-Bi2212 heterostructure to become non-superconducting. We achieved zero resistance in a two-unit-cell-thick Bi2212 (TUC-Bi2212)-based heterostructure as expected (Fig. 3a&b). However, after oxidation, the temperature range associated with the transition was more than twice that observed before oxidation, and $T_c$ was decreased slightly (Fig. 3d). Graphene/SUC-Bi2212 and graphene/TUC-Bi2212 heterostructures fabricated together were measured before oxidation, and both were superconducting. After oxidation, TUC-Bi2212 remained superconductive and reached zero resistance (Fig. 3a), whereas SUC-Bi2212 lost its superconductivity. This was not a result of destruction of graphene. If the graphene/SUC-Bi2212 sample had been damaged and had lost its superconductivity, the resistance would have gone up to several hundred mega-ohms based on our investigations. In our case, the resistance of SUC-Bi2212 was only several kilo-ohms after oxidation. This suggested that SUC-Bi2212 was not destroyed completely by oxidation.

**Investigation on the normal state with reduced thickness**

The preservation of superconductivity in ultrathin films provides a rare opportunity to explore insights on the nature of superconductivity. In particular, it allows us to test whether correlations between certain transport properties and superconductivity are important to the pairing mechanism, as previously suggested[23].

It is well known that the resistivity of cuprates around optimal doping increases linearly with increasing temperature over a wide range (often described as $T$-linear $\rho$)[30]. We found that the resistance of Bi2212 in the normal state increased linearly with increasing temperature not only at 270-unit-cell thickness (which can be considered a bulk material, because its $R$-$T$ curve is the same as that of the bulk[31]), but even down to half-unit-cell thickness (Fig. 4a). In addition, the slope decreases symmetrically with reduced thickness (Fig. 4b).

The slope of this linear resistivity ($A$), which represents the strength of the scattering, scales with $T_c$ in bulk cuprate, organic, and pnictide superconductors[23] (as demonstrated by the pink squares, olive circles, and blue triangles in Fig. 4c, respectively). However, the empirical correlation between $A$ and $T_c$ fails in the graphene/ultrathin Bi2212 heterostructure (black stars in Fig. 4c), in which $A$ changes by a factor of 4-5, whereas $T_c$ varies by only about 10%. We note that the range of $A$ values in our data is much larger than that presented in previous data, in which $A$ remained very small, demonstrating the robustness of our conclusions.

Although the sheet resistance ($R_\square$) in the normal state of Bi2212 increased from ohms to kilo-ohms as the thickness was reduced from 270 unit cells to 0.5 unit cells, $T_c$ remained almost the same (Fig.4a&d). In contrast, the result from ultrathin amorphous Mo-Ge films (blue diamonds in Fig. 4d) showed that $T_c$ decreases rapidly with increasing sheet resistance[32], which is attributed to disorder induced Anderson localization and increased Coulomb interaction[33]. This discrepancy suggests that the increase in $R_\square$ with reduced thickness in Bi2212 is unrelated to disorder. Instead, this increase may be related to a decrease in the number of conducting paths through inter-plane coupling between $CuO_2$ layers, a process unrelated to $T_c$.

**Discussion**

Robust superconductivity has already been reported in some atomically thin conventional superconductors[34-36]. Our investigations reveal that this phenomenon could also be observed in high-$T_c$ superconductor. Superconductivity can be preserved in ultrathin Bi2212 even at half-unit-cell thickness, with the protection of graphene. The *I-V* behavior from the combined graphene/SUC-Bi2212 structure may inspire the development of new superconducting devices that benefit from both the robust electronic properties of graphene and the high speed of superconductivity.

In addition to graphene being crucial for the preservation of superconductivity in ultrathin Bi2212, the normal state behaviors of Bi2212 with reduced thickness are also essential to understanding high-$T_c$ superconductivity. Normal state transport in cuprates is considered to be strongly related to the pairing mechanism[23]. Surprisingly, in a sample of graphene/ultrathin Bi2212 heterostructure the normal state properties were decoupled from the superconductivity. The disconnection of *A* and $T_c$ brings into question the link between the large angle scattering process and the superconductivity mechanism, which suggests the need to search for other avenues to further understand the origin of high-$T_c$ superconductivity in cuprates. Therefore, graphene/ultrathin Bi2212 heterostructure samples provide a promising platform to obtain further insight into the underlying mechanisms of superconductivity and for the exploration of new devices.

## Methods

**Exfoliation of Bi2212**

A 5 mm-wide and 0.5 mm-thick platelet of single-crystal Bi2212 was exfoliated using scotch tape and transferred onto a $SiO_2$/Si substrate ($SiO_2$-300 nm; Si-500 μm)[37]. Bi2212 always cleaves between two BiO planes, in which Van der Waals-type bonding is present[38]. To select ultrathin Bi2212 flakes from the exfoliated flakes on the substrate, a combination of optical microscope and atomic force microscope was used[27].

**Transfer of chemical vapour deposition (CVD) graphene**

Continuous graphene films with domains larger than 100×100 μm$^2$ were grown on 1×1 cm$^2$, 25 μm thick polished copper foil (Alfa Aesar) using CVD[39]. The as-grown graphene on Cu was spin-coated with poly-methyl methacrylate (PMMA) and then covered with a piece of tape. Ammonium persulphate (0.1 M) was used to dissolve the copper foil[39]. The tape-PMMA-graphene was then dry transferred on top of the substrate with Bi2212 flakes immediately after the flakes were cleaved. To obtain a clean surface and better contacts between graphene and Bi2212, thermal annealing was performed in $O_2$/Ar (20 sccm/200 sccm) at 200°C for 30 minutes in a tube furnace[40] after removing the PMMA and the tape with acetone. Contamination at the interface of the heterostructure can clean itself[41-43].

**Device fabrication**

The microfabrication process that has been applied for graphene-based devices[27], was used to prepare devices from a graphene/Bi2212 van der Waals heterostructure. A mask aligner was used to create an electrode pattern on top of the heterostructures. Then, a layer of 50 nm Au/5 nm Cr was deposited by electron-beam deposition followed by standard lift-off procedures.

After device fabrication and initial transport measurements, the heterostructure was exposed to ultraviolet light in ozone[29] to oxidize the graphene protection layer and achieve zero resistance in the superconducting state.

40. Britnell, L. *et al.* Field-effect tunneling transistor based on vertical graphene heterostructures. *Science* **335**, 1140-1143(2012).

41. Haigh, S. J. *et al.* Cross-sectional imaging of individual layers and buried interfaces of graphene-based heterostructures and superlattices. *Nat. Mater.* **11**, 764-767 (2012).

42. Georgiou, T. *et al.* Vertical field-effect transistor based on graphene $WS_2$ heterostructures for flexible and transparent electronics. *Nat. Nanotechnol.* **8**, 100-103 (2013).

43. Lui, C. H., Liu, L., Mak, K. F., Flynn, G. W. & Heinz, T. F. Ultraflat graphene. *Nature* **462**, 339-341 (2009).
**Acknowledgments**

We sincerely thank Z. X. Shen, D.H. Lee, A.K. Geim, S.Q. Shen for discussions during the preparation of this manuscript.

This work was partially supported by the Strategic Priority Research Program (B) of the Chinese Academy of Sciences (Grant No. XDB04010400, XDB04040300 and XDB04020000）, National Science and Technology Major Project (Grant No. 2011ZX02707), the National Natural Science Foundation of China (Grant No. 11104303, 11274333, 11204339, 11104335, and 61136005), Chinese Academy of Sciences (Grant No.KGZD-EW-303KJCX2-EW-W11, and XDA02040000), Hundred Talents Program of CAS, CAS International Collaboration and Innovation Program on High Mobility Materials Engineering, Science and Technology
15

Commission of Shanghai Municipality (Grant No. 12JC1410100 and 12JC1403900), and the Ministry of Science and Technology of China (Grant No. 2011CBA00107 and 2012CB921302).

## Author contributions

D.J. planned the research and produced the samples. D.J. and T.H. designed the experiments. T.H. And Q.L. performed the measurements. T.H. analyzed the data. Z.C., H.Z., and G.Y. grew graphene by CVD. C.L. grew the Bi2212 single crystals. D.J., Q.S., and J.Z. performed the micro-fabrication. D.J., T.H., L.Y., A.L., H.W., G.M., H.X., X.X., and M.J. discussed the results. D.J. and T.H. wrote the paper. M.J. and X.X. supervised the work.

## Additional Information

**Competing financial interests:** The authors declare no competing financial interests.

## Figure Legends

**Fig. 1. Graphene/Bi2212 heterostructure.** (**a**) Optical-microscope image (in white light) of several Bi2212 flakes included HUC-Bi2212 and SUC-Bi2212 (see the Methods section for the detailed fabrication procedure). Colors: light purple--$SiO_2$/Si substrate; blue--SUC-Bi2212 flake; light blue--the thick Bi2212 flake. Scale bar: 20 μm. (**b**) Atomic force microscope image of a Bi2212 flake. Color: dark yellow--HUC-Bi2212; bright yellow--SUC-Bi2212; brown--

SiO$_2$/Si substrate. The inset shows the AFM profile of the flake. The lower step is approximately 2.7 nm and the upper step is approximately 2.3 nm. However, one should consider that the thickness measured by AFM includes the chemical and van der Waals contrasts[24] and the distance between Bi2212 and SiO$_2$, so that the measured thickness is always larger than the theoretical values and the differences vary from sample to sample with a typical value of around 1 nm. The theoretical thicknesses of HUC-Bi2212 and SUC-Bi2212 are 1.5 nm and 3 nm, respectively, so that for thicknesses of 2.7 nm and 2.3 nm, they can only be half-unit-cell-thick. Scale bar: 5 μm. (**c**) Schematic view of the device. The SUC-Bi2212 flakes were prepared by micromechanical exfoliation of a bulk Bi2212 single crystal, which cleaves between two BiO planes, and for which van der Waals-type[25] bonding is present. Each unit cell consists of two CuO$_2$-bilayer blocks sandwiched in between Sr-Bi-O charge reservoir layer. Single-layer graphene prepared by using chemical vapour deposition (CVD) was then transferred on top of the SUC-Bi2212 flakes. The electrodes were prepared using microfabrication techniques. (**d**) An optical image of one of our experimental devices prepared from a graphene/Bi2212 van der Waals heterostructure sample. Scale bar: 10 μm. u.c. = unit cell.

**Fig. 2. Typical changes in temperature dependence with differential resistance.** (**a**) *R-T* curves for graphene/SUC-Bi2212 van der Waals heterostructures measured at applied bias currents of $I_{dc}$= 0, 0.04, 0.08, 0.16, 0.20, 0.24, 0.28, 0.32, and 0.36 mA. The transition temperature ($T_c$) of each thickness was defined as the intersection of the two straight lines. Inset:

brief circuit diagram of the transport measurement. (**b**) *R-T* curves for graphene/SUC-Bi2212 heterostructures with different bias currents. $I_{dc}$ = 0.16, 0.20, 0.24, 0.28, 0.32 and 0.36 mA, after subtraction of the differential resistance for the interface between graphene and SUC-Bi2212. (**c**) Temperature dependence of normalized resistance curves for three graphene/SUC-Bi2212 heterostructure samples after subtraction of the temperature independent residual sheet resistance (typically several ohms): Sample 1 (black line), Sample 2 (olive line), and Sample 3 (blue line). These three curves are normalized at 250 K and can be scaled with each other. (**d**) *I-V* curves for graphene/SUC-Bi2212 van der Waals heterostructure measured at 10 K ≤ *T* ≤ 300 K.

**Fig. 3. Comparison between graphene and graphene-oxide protection layers.** (**a**) *R-T* curves for a graphene/TUC-Bi2212 van der Waals heterostructure sample before and after oxidation of the graphene protection layer. The orange curve represents the sample with the graphene protection layer, black curve represents the same sample after oxidation of the graphene protection layer. (**b**) Normalized (250 K) *R-T* curves of (**a**). (**c**) Illustration of the device before and after oxidation. The active area decreases after oxidation as a result of etching by oxygen, which partially contributes to the change in the resistance in the normal state. (**d**) Expanded version of the data in (**b**). The transition width is defined as the range between $T_c$ and the temperature with 10% resistance of $T_c$'s.

**Fig. 4. *R-T* behavior of ultrathin Bi2212.** (**a**) *R-T* curves for Bi2212 with various thicknesses

from 270-unit-cell-thick to half-unit-cell-thick. (**b**) Temperature dependence of normalized (250 K) sheet resistance of Bi2212 with thickness ranging from 270 unit cells to a half unit cell. The solid lines represent samples in this investigation. The dark cyan diamond dots represent the bulk Bi2212 in Reference 31. (**c**) $A$ versus $T_c/T_c^{max}$. $A$ is the slope for $T$-linear $\rho$ in the normal state. $T_c^{max}$ is $T_c$ of the optimally doped bulk sample. $T_c^{max}$ is 93 K for Bi2212, 90 K for $Tl_2Ba_2CuO_{6+\delta}$, 1.1 K for $(TMTSF)_2PF_6$, and 26 K for $Ba(Fe_{1-x}Co_x)_2As_2$. The black stars represent graphene/ultrathin Bi2212, the pink squares represent $Tl_2Ba_2CuO_{6+\delta}$, the olive circles represent $(TMTSF)_2PF_6$, and the blue triangles represent $Ba(Fe_{1-x}Co_x)_2As_2$ in Ref. 23. (**d**) Sheet resistance in the normal state versus $T_c$. The black stars are taken from Fig. 4a and the blue diamonds represent $Mo_{79}Ge_{21}$, acquired from Ref. 32. $T_c^{max}$ is 7.2 K for $Mo_{79}Ge_{21}$. u.c. = unit cell.

**Fig. 1**

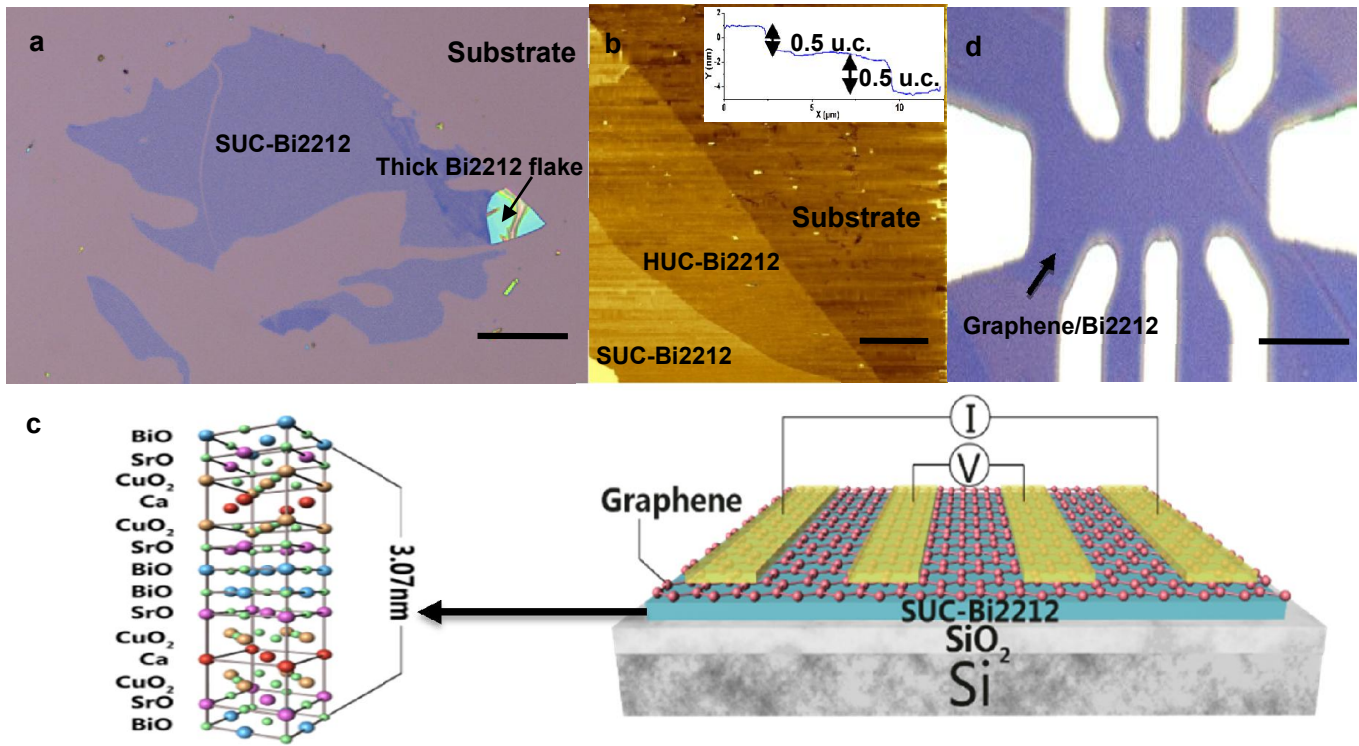

**Fig. 2**

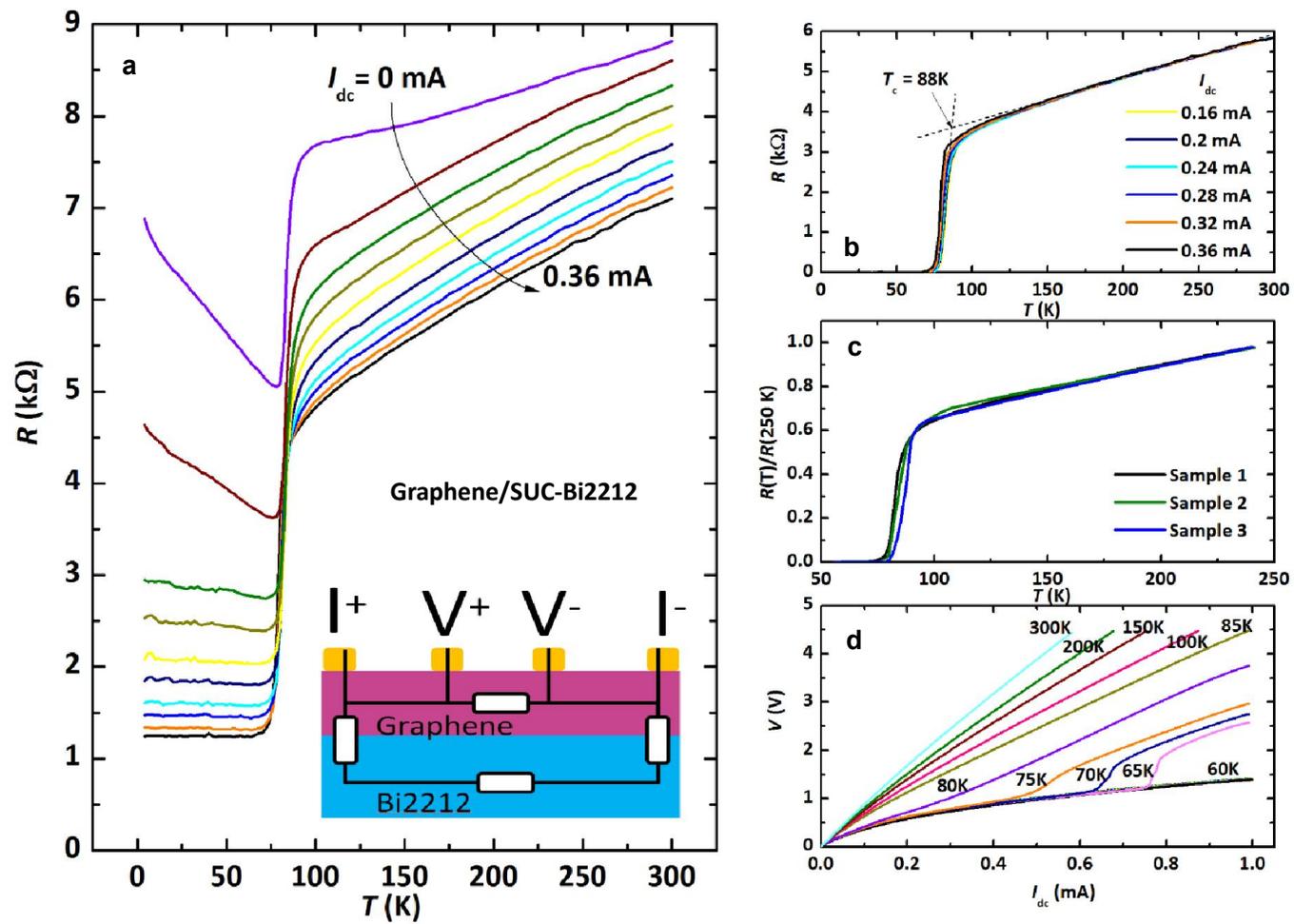

**Fig. 3**

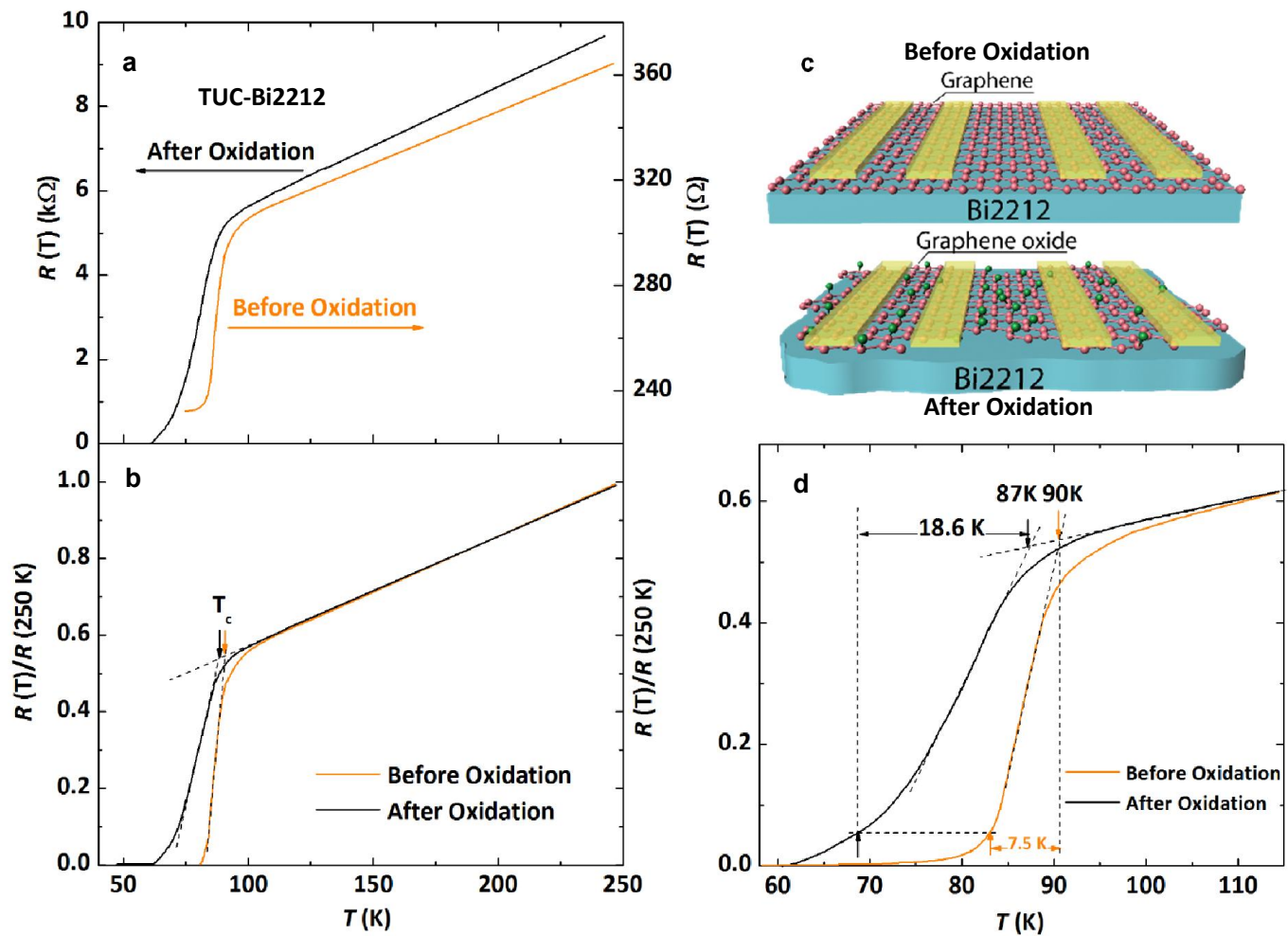

**Fig. 4**

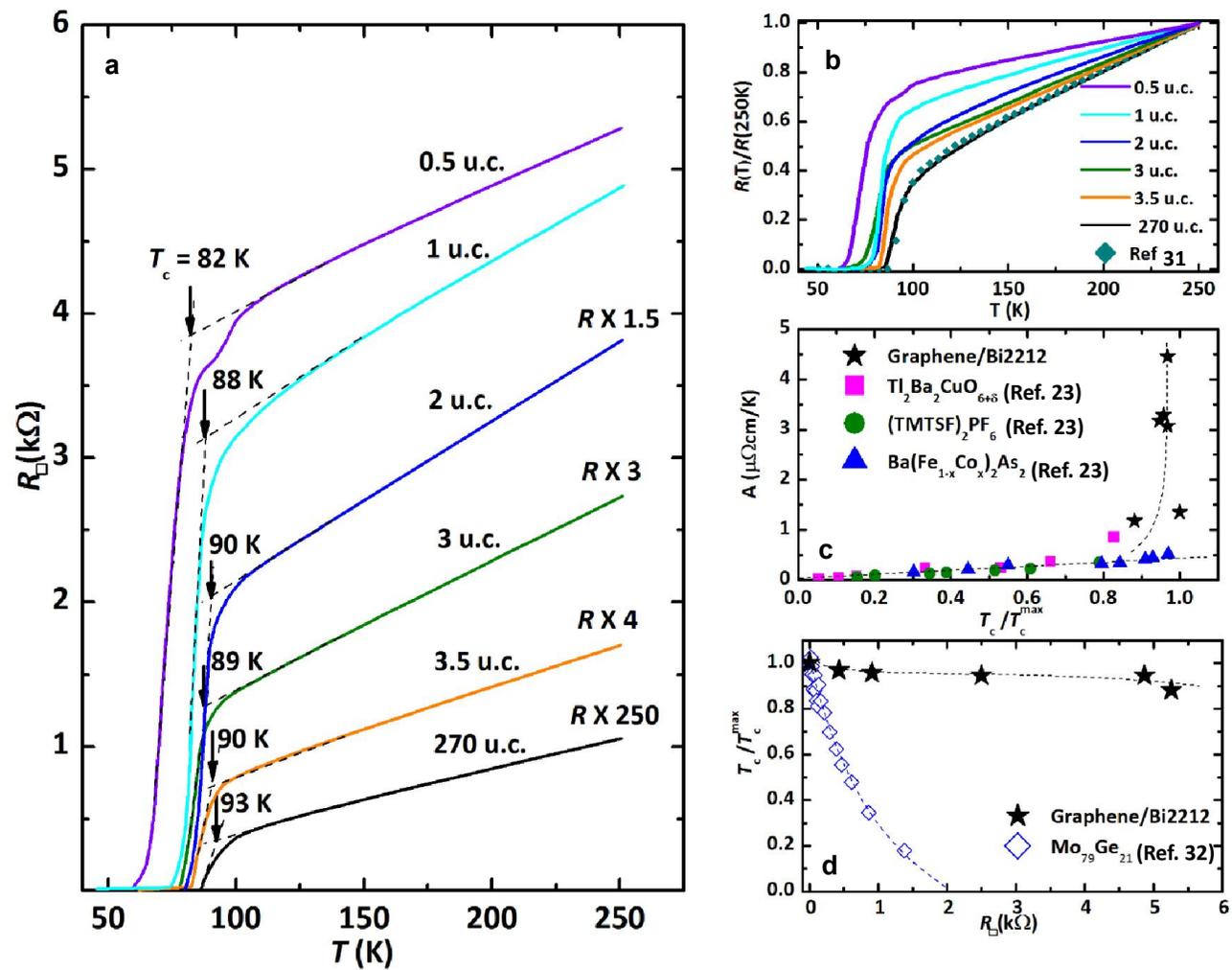

23